\documentclass[lettersize, journal]{IEEEtran}

\usepackage{cite, hyperref}
\usepackage{amsmath,amssymb,amsfonts}
\usepackage{ulem}
\usepackage{graphicx}
\usepackage{textcomp}
\usepackage{algorithm, algpseudocode}
\usepackage{xcolor}
\usepackage{nicefrac, diagbox}
\usepackage[font=footnotesize, skip=2pt]{caption}
\usepackage{subcaption}
\begin{document}

\title{Generalized Spectral Clustering of Low-Inertia Power Networks}

\author{Gerald~Ogbonna,~\IEEEmembership{Student~Member,~IEEE,}
        and~C. L.~Anderson,~\IEEEmembership{Senior~Member,~IEEE}
\thanks{G. Ogbonna is with the Department of Systems Engineering, Cornell University, Ithaca, NY, 14853 USA (e-mail: gco27@cornell.edu).}
\thanks{C. L. Anderson is with the Department of Biological and Environmental Engineering, Systems Engineering, and the Center for Applied Mathematics, Cornell University, Ithaca, NY, 14853 USA (email: cla28@cornell.edu).}
\thanks{Authors would like to thank the Cornell Atkinson Center for Sustainability and the National Science Foundation for support of this work.\\ Manuscript received XX, 2025; revised XX}
}

\maketitle

\begin{abstract}
Large-scale integration of distributed energy resources has led to a rapid increase in the number of controllable devices and a significant change in system dynamics. This has necessitating the shift towards more distributed and scalable control strategies to manage the increasing system complexity. In this work, we address the problem of partitioning a low-inertia power network into dynamically coherent subsystems to facilitate the utilization of distributed control schemes. We show that an embedding of the power network using the spectrum of the linearized synchronization dynamics matrix results in a natural decomposition of the network. 
We establish the connection between our approach and the broader framework of spectral clustering using the Laplacian matrix of the admittance network. The proposed method is demonstrated on the IEEE 30-bus test system. We consider the robustness of the clusters by analyzing the sensitivity of the small eigenvalues and their corresponding eigenspaces to perturbations caused by variation in the steady-state operating points of the network.

\end{abstract}


\section{Introduction}
\IEEEPARstart{T}{he} transition to a more sustainable energy system typically requires the large-scale integration of renewable generation resources, micro-generators, energy storage devices, and responsive loads, engaged in real-time balancing of supply and demand across different levels of the grid. Recent regulatory changes, such as FERC order $2222$ \cite{noauthor_ferc_nodate}, enable direct participation of DERs in electricity markets, ensuring the grid of the future will be composed of a large number of these controllable resources that require real-time coordination. In addition, the gradual replacement of synchronous generators with intermittent and variable renewable generation leads to a gradual loss of the robust frequency and voltage control inherent to synchronous machines\cite{dorfler_distributed_2019}. This increasingly complex system requires new operational paradigms, including optimization-based control schemes for automatic generation control \cite{venkat_distributed_2008}, which have been shown to be fast-acting and spatially precise, enabling coordinated response to system disturbances.

In most modern power systems, the independent system operator (ISO) is responsible for the centralized coordination of all system resources. However, as the number of decision variables and operational constraints increases, solving the real-time centralized coordination and control problem can become computationally prohibitive. Furthermore, the communication requirement between the centralized controller and these resources poses an additional scalability challenge.

Several works \cite{yazdanian_distributed_2014, huo_distributed_2021, chanfreut_survey_2021} have shown the potential of distributed control schemes -- such as distributed model-predictive control\cite{venkat_distributed_2008, venkat_distributed_2006}, multi-agent systems\cite{lian_game-theoretic_2017, chanfreut_fast_2022}, and consensus-based methods\cite{cady_distributed_nodate, dominguez-garcia_distributed_2011} for achieving acceptable trade-offs between algorithmic performance and communication/coordination requirements for centralized control problems. Chanfreut et al. \cite{chanfreut_survey_2021} further showed that balancing the dual objectives of performance and communication requirements in distributed control schemes is analogous to that of system partitioning, which seeks to identify a suitable decomposition of a global system into suitable subsystems. In the context of networked control systems, this decomposition can be framed as a network partitioning problem, where the partitions should be such that the control actions and disturbances originating within each subsystem are reasonably confined to that partition, with minimal impact on the rest of the network. In this work, the term \textit{partition} is used synonymously with cluster, control zone, subsystem, and control area.

In conventional power networks (such as PJM), the current zones and zonal boundaries are determined by historical asset ownership rather than by the properties of the network \cite{cotilla-sanchez_multi-attribute_2013}. Control zones defined in this way may be sufficient for system planning and markets, but are not necessarily suited for decentralized power system operations. To address this need, recent works have approached network partitioning in power networks with different objectives including contingency management \cite{sun_splitting_2003, ding_two-step_2013}, generator coherency identification \cite{romeres_novel_2013}, designing localized reactive power markets \cite{zhong_localized_2004}, localized voltage control \cite{lagonotte_structural_1989, nayeripour_coordinated_2016}, and parallel power system restoration \cite{ganganath_agglomerative_2018} using methods including spectral clustering techniques \cite{ng_spectral_2001, sanchez-garcia_hierarchical_2014, ding_two-step_2013, wu_spectral_2023}, evolutionary computation algorithms \cite{ding_optimal_1994, 
blumsack_defining_2009}, information theoretic approaches \cite{lagonotte_structural_1989, baranwal_clustering_2017}, and sensitivity analysis techniques \cite{tamp_sensitivity_2014, 
nayeripour_coordinated_2016
}.

In this work, we address the problem of partitioning power networks into control areas to facilitate scalable distributed control strategies. Some of the limitations of previous approaches \cite{sanchez-garcia_hierarchical_2014, tyuryukanov_graph_2020} are that they use a DC power flow model of the network, this approximation may fail to capture the true underlying network dynamics, especially when the phase angle differences become large. Additionally, many classic spectral clustering approaches for graphs \cite{ng_spectral_2001} do not typically take into account node types, consequently, using them for clustering in power network can result in clusters without generator nodes. We extend the spectral clustering approach for power networks developed in \cite{sanchez-garcia_hierarchical_2014}, where the spectrum of the Laplacian of a graph whose edge weights represent branch susceptances is used to identify clusters in the network.


The main contributions of this paper are as follows:
\begin{enumerate}
    \item We develop a generalized spectral clustering framework based on the linearized synchronization dynamics of low-inertia power networks, providing a dynamically meaningful decomposition of the system. 
    \item We formulate the clustering problem as a generalized eigenvalue problem that incorporates both network structure and heterogeneous damping, enabling the method to capture the influence of device-level dynamics on system behavior. 
    \item We provide conditions under which the identified clusters are robust to variations in the steady-state network operating condition.
    \item We demonstrate this approach on the IEEE 30-bus network for various numbers of clusters, showing that the resulting clusters are dynamically coherent with perturbations largely confined to the originating cluster.
    \item We provide a meaningful interpretation of the resulting clusters in the context of power systems.
\end{enumerate}

The remainder of the paper is organized as follows: Section \ref{sec: preliminaries} presents graph-theoretic and power-system preliminaries; Section \ref{sec: clustering} describes the clustering optimization problem and the generalized spectral clustering solution. Section \ref{sec: results} presents results for a test power system, with simulations of the impact of disturbances on network dynamics and a robustness analyses of the resulting clusters presented in Section \ref{sec: numerical_validation}. Finally, Section \ref{sec: conclusion} concludes the paper.

\section{Preliminaries}
\label{sec: preliminaries}
\subsection{Graph Theory Preliminaries}
Given a symmetric matrix $A$, the notation $A \succ 0$ and $A \succeq 0$ denotes positive definiteness and semi-definiteness, respectively. $\mathbf{1}_n \in \mathbb{R}^{n}$ and $\mathbf{0}_{n} \in \mathbb{R}^n$ denotes the $n$-dimensional vectors of all ones and zeros, respectively.

We denote by $\mathcal{G}$ the undirected graph defined as the triple $\mathcal{G} = (\mathcal{V}, \mathcal{E}, W)$, $\mathcal{V} = \{1, \ldots , n \}$ is the set of vertices and $|\mathcal{V}| = n$ is the number of vertices on the graph, $\mathcal{E} \subseteq \mathcal{V} \times \mathcal{V}$ denotes the edge set (i.e. branches) on the graph with $|\mathcal{E}| = m$ edges, and $W \in \mathbb{R}^{n \times n}$ denotes the weighted adjacency matrix. We use $\{i,j\} \in \mathcal{E}$ to represent the undirected edge connecting nodes $i, j \in \mathcal{V}$ with edge weight $w_{ij}\geq 0$. For a non-empty set $\mathcal{V}_1 \subset \mathcal{V}$, we denote the boundary of the set $\mathcal{V}_1$ as $\partial(\mathcal{V}_1) = \sum_{i \in \mathcal{V}_1, \, j \in \overline{\mathcal{V}}_1} w_{ij}.$

In this work, we assume the graph $\mathcal{G}$ is simple, that is, $w_{ii} = 0$ for all $i$ and a single edge connects every pair of connected nodes. 
For each branch $\{i,j\} \in \mathcal{E}$, if we assign an arbitrary orientation and a unique number $\ell \in \{1, \ldots, m\}$, the oriented incidence matrix of the denoted by $B \in \mathbb{R}^{n \times m}$ is defined elementwise as $B_{k \ell} = 1$ if node $k$ is a sink of the oriented edge $\ell$, $-1$ if node $k$ is a source for edge $\ell$, and $0$ otherwise.

The Laplacian matrix of the graph denoted by $L \in \mathbb{R}^{n \times n}$ is the symmetric positive semi-definite matrix defined as $L = D - W$, where $D$ is the diagonal matrix of node degrees $D_{ii} = \sum_{j} w_{ij}$ for each $i$. If we denote the $i$th smallest eigenvalue of $L$ by $\lambda_i$, the eigenvalues can be ordered as $0 = \lambda_{1} \leq \lambda_2 \leq \ldots \leq \lambda_{n}$. If the graph $\mathcal{G}$ is a connected, the second smallest eigenvalue (the algebraic connectivity) $\lambda_{2} > 0$ and $\text{Ker}(L) = \text{Ker}(B^{\intercal}) = \text{Span}(\mathbf{1}_{n})$, that is $L\mathbf{1}_n = \mathbf{0}_{n}$, this subspace is commonly referred to as the agreement subspace of $\mathcal{G}$ \cite{ainsworth_structure-preserving_2013}.

Given a pair of symmetric matrices $L, D \in \mathbb{R}^{n \times n}$, we denote by $\lambda_i(L, D)$ the $i$th smallest generalized eigenvalue of the matrix pair $(L, D)$. $\lambda_i(L, D) \in \mathbb{R}$ is the scalar that solves the generalized eigenvalue equation $Lv = \lambda_i(L, D)Dv$, and the non-zero vector $v \in \mathbb{R}^n$ is the corresponding eigenvector.

\subsection{Power System Preliminaries}
In this work, we consider a lossless AC power system, that is, $r_{ij} = 0$ for all $i$ and $j$. Ignoring the effect of line charging (shunts) and assuming that the 3-phase system is balanced, the network can be modeled as an undirected weighted graph $\mathcal{G} = (\mathcal{V}, \mathcal{E}, W)$, with $|\mathcal{V}| = n$ nodes (buses) and $|\mathcal{E}| = m$ branches (transmission lines). We assume that the set of nodes can be partitioned as $\mathcal{V} = \mathcal{V}_G \cup \mathcal{V}_R \cup \mathcal{V}_L$, where $\mathcal{V}_G$ is the set of buses with synchronous generators, $\mathcal{V}_R$ is the set of buses with renewable generation connected to the grid via droop controlled inverters, and $\mathcal{V}_L$ is the set of load buses. The edge weight of the branch $\{i,j\} \in \mathcal{E}$ is given by $w_{ij} = \Im(Y_{ij})$, where $Y = Y^{\intercal} \in \mathbb{C}^{n \times n}$ is the complex admittance matrix (in per unit) of the network.

We denote the per unit bus voltage magnitude at bus $i$ by $|V_{i}| \in \mathbb{R}$ and the voltage angle (in rad) as $\delta_{i} \in \mathbb{S}^{1}$, where $\mathbb{S}^{1}$ is the unit circle. Given $\delta \in \mathbb{T}^n = \mathbb{S}^{1} \times \mathbb{S}^{1}\times \cdots \times \mathbb{S}^{1}$ -- the vector of bus voltage phase angles in the $n$-torus -- we define the vector of line angles as $\theta = B^{\intercal}\delta \in \mathbb{T}^{m}$, where $\theta_{\ell} = \delta_i - \delta_j$ for each edge $\ell \in \mathcal{E}$. The real power flow on line $\ell \in \mathcal{E}$ can be expressed in terms of the bus voltage magnitudes, angles, and the entries of the admittance matrix as 
\begin{align*}
    p_{ij}&=|V_{i}||V_{j}|\Im(Y_{ij})\sin(\theta_{\ell})
\end{align*}
where $\theta_l = \delta_i - \delta_j$ is the line angle across the transmission line $\ell$. Dörfler et al \cite{dorfler_synchronization_2013, dorfler_exploring_2012} showed in  that 
the dynamics of real power flow in a lossless power network can be characterized by the generalized coupled oscillator model
\begin{align}
    \label{eq: 1} M_{i} \ddot{\delta}_{i} + D_{i} \dot{\delta}_{i} &= \omega_{i} - \sum\limits_{j = 1}^{n} p_{ij}, \; i \in \mathcal{V}_{G} \\
     \label{eq: 2} D_{i}\dot{\delta}_{i} &= \omega_{i} - \sum\limits_{j = 1}^{n} p_{ij}, \; i \in \mathcal{V}_{L} \cup \mathcal{V}_{R}
\end{align}
where $M_{i} > 0$ is the inertia coefficient of the synchronous generator connected at node $i \in \mathcal{V}_{G}$. The term $D_i > 0$ represents the damping coefficient of the $i$th synchronous generator when $i \in \mathcal{V}_{G}$, the frequency-dependent component of the load connected to bus $i$ when $i \in \mathcal{V}_{L}$, and for droop-controlled inverter buses $i \in \mathcal{V}_{R}$, the total frequency dependence from the inverter's droop and frequency-responsive loads (i.e., $R_i^{-1} + D_i'$, where $R_i^{-1}$ is the droop coefficient and $D_i'$ is the frequency-responsive component of the load) \cite{ainsworth_structure-preserving_2013, dorfler_exploring_2012}, and $\omega_{i}$ is the natural frequency of the $i$th oscillator (this is the net real power injection at bus $i$).

With the integration of more renewable generation and the retirement of synchronous generation, we assume that the inertia coefficients $M_{i}$ are small compare damping term $D_i$ -- this is typical for low inertia grids. \cite{dorfler_synchronization_2012} showed that under these assumptions, in the time-scale for synchronization, the network dynamics can be approximated by the dynamics of non-uniform Kuramoto oscillators where the model's approximation error is $\mathcal{O}(\epsilon)$ and goes to zero as $t \rightarrow \infty$, $\epsilon>0$ is the singular perturbation parameter which typically is the worst-case choice of the ratio $M_{i}/D_{i}$. The slower dynamics of the network, that approximates the synchronization dynamics, is given by the reduced first-order model
\begin{align}
    \label{eq: 3} D_{i}\dot{\delta}_{i} &= \omega_{i} - \sum\limits_{j = 1}^{n} |V_{i}||V_{j}|\Im(Y_{ij}) \sin(\delta_{i} - \delta_{j}), \; i \in \mathcal{V}
\end{align}
where $D_i$ in (\ref{eq: 3}) is the \textit{effective damping} (time constant) of the $i$th first-order oscillator and accounts for the total synchronous generator's damping (including the effect of internal excitation control), droop coefficients of an inverter, and the frequency dependence of loads.

\section{Clustering}
\label{sec: clustering}
\subsection{Linearized Synchronization Dynamics}
\label{subsec: linearized_dynamics}
Let $|V^{*}|$ and $\delta^{*}$ denote the voltage magnitudes and phase angles at a frequency-synchronized solution (i.e. a power flow solution),
if we assume that reactive power balance in the network can be controlled via local compensation, meaning that the voltage magnitudes $|V_i^*|$ can be assumed to be constant (not necessarily $1.0$ p.u.), the linearized voltage angle dynamics around $\delta^*$ is given by the weighted Laplacian flow,
\begin{align}
    \label{eqn: state_dynamics} D\dot{\Delta\delta} = -\tilde{L}\Delta\delta
\end{align}
where $\Delta \delta = \delta - \delta^*$ is the deviation in the voltage angle from $\delta^*$, $D$ is the diagonal matrix of non-negative effective damping coefficients, and $\tilde{L}$ is the Laplacian matrix of the graph whose edge weights are given by $\tilde{w}_{ij} = |V_{i}^{*}||V_{j}^{*}|\Im(Y_{ij})\cos{(\delta_i^* - \delta_j^*)}$, the sensitivity of the real power flow $p_{\ell}$ on line $\ell = \{i,j\}$ to the line angle $\theta_{\ell}$ around the synchronized solution. We refer the reader to the work \cite{ogbonna_structure_2026} for details of this derivation. The undirected weighted graph $\tilde{\mathcal{G}} = (\mathcal{V}, \mathcal{E}, \tilde{W})$ whose Laplacian matrix is $\tilde{L}$ is often referred to as the \textit{Dynamic Graph} of the power network \cite{ding_two-step_2013} with edge weights $\tilde{w}_{ij}$ (commonly referred to as the \textit{synchronizing coefficients} \cite{musca_wide-area_2025}) of the graph $\{i, j\}$ and represents the strength of coupling between the oscillators at nodes $i$ and $j$. The time constants $D_i > 0$ in (\ref{eqn: state_dynamics}) represent the node weights on the graph $\tilde{\mathcal{G}}$.

Notice that for a predominately inductive network, the edge weights $\tilde{w}_{ij} \geq 0$ for all $\{i,j\} \in \mathcal{E}$ when the phase angles $\delta^* \in \overline{\Delta} = \{\delta\; | \;||B^{\intercal}\delta||_{\infty} < \frac{\pi}{2} \text{ rad} \} \subset \mathbb{T}^n$, and the Laplacian matrix $\tilde{L}$ is positive semi-definite. This is a generalization of the results in ~\cite{sanchez-garcia_hierarchical_2014} which uses a DC power flow approximation (reducing the coupling terms $\tilde{w}_{ij}$ to $\Im (Y_{ij})$) of the network for clustering. As the steady-state line angle $\theta_{\ell}^*$, which is shown in \cite{dobson_combining_2010} to be a measure of the stress across the transmission line $\ell$, approaches $\frac{\pi}{2}$ rad, $\cos(\theta_{\ell}^*) \rightarrow0$, the edge weights on the dynamic graph $\tilde{w}_{ij}\rightarrow 0$.

Since the diagonal matrix of time constants $D$ is non-singular, we can rewrite the linearized synchronization dynamics in state-space form as
\begin{align*}
    \dot{\Delta\delta} = -D^{-1}\tilde{L}\Delta\delta = J\Delta \delta,
\end{align*}
where $J$ is the system dynamics matrix. $J$ is similar to the symmetric matrix $-D^{-1/2}\tilde{L}D^{-1/2}$ via the transformation matrix $P = D^{-1/2}$, hence they have the same real eigenvalues (the negative of the eigenvalues of the pair $(\tilde{L}, D)$) and if $u$ is an eigenvector of $-D^{-1/2}\tilde{L}D^{-1/2}$, then $v = Pu = D^{-1/2} u$ is the corresponding eigenvector of $J$ (the generalized eigenvector of the pair $(\tilde{L}, D)$). For a connected network with voltage angles $\delta \in \overline{\Delta}$, the zero eigenvalue of $J$ has an algebraic multiplicity of $1$ and the corresponding eigenvector is $\mathbf{1}_n$.

For a network of oscillators with identical time constants (i.e., $D = \alpha I$ for some $\alpha>0$), the $i$th eigenvalue of the system dynamics matrix $J$ reduces to the scaled eigenvalues of $\tilde{L}$, that is $\lambda_i(J) = -\lambda_i(\tilde{L})/\alpha$, while the eigenvectors are exactly the eigenvectors of $\tilde{L}$. In the general case of a network with non-uniform time constants, the Laplacian matrix $\tilde{L}$ and the diagonal matrix of time constants $D$ play a crucial role in determining the evolution of the state variables and the community structure of the network. We show in Section \ref{subsec: optimization_problem} that the eigenvectors corresponding to the smallest eigenvalues of $J$ (also referred to as the inter-area modes of the system) can be used to decompose the network into groups  of dynamically coherent nodes.

\subsection{The Clustering Problem}
\label{subsec: optimization_problem}
The clustering problem on the network, can be defined as the problem of finding $2  \leq k \ll n$ sets $\mathcal{S}_1, ,\ldots, \mathcal{S}_k \subset \mathcal{V}$ such that $\mathcal{S}_i \neq \mathcal{V}$ or $\emptyset$, for all $i \neq j$ $\mathcal{S}_{i} \cap \mathcal{S}_{j} = \emptyset$, and $\bigcup_{i = 1}^{k} \mathcal{S}_{i} = \mathcal{V}$, so that nodes within each set $\mathcal{S}_{i}$ are strongly coupled (hence form coherent groups), while being loosely coupled to the rest of the nodes $\mathcal{S}_{i}^{c}$ in the network. Clusters defined in this way ensure that network disturbances or control actions originating within each cluster are reasonably confined to that cluster.

For each set $\mathcal{S}_{i} \subset \mathcal{V}$, the characteristic vector $\chi_{\mathcal{S}_i}$ (also the cut vector of the induced subgraph) is defined element-wise as $[\chi_{\mathcal{S}_i}]_k = 1$ if $k \in \mathcal{S}_{i}$ and $0$ otherwise. For each candidate partition $\mathcal{S}_{i}$, we define a measure of \textit{badness} in terms of the Laplacian matrix of the dynamic graph $\tilde{L}$ and the diagonal matrix of effective damping $D$ as
\begin{align*}
    \mathbf{\phi}(\mathcal{S}_{i}) = \frac{\chi_{\mathcal{S}_i}^{\intercal} \tilde{L}\chi_{\mathcal{S}_i}}{\chi_{\mathcal{S}_i}^{\intercal} D \chi_{\mathcal{S}_i}}
\end{align*}
where the quadratic form 
\begin{align*}
    \chi_{\mathcal{S}_i}^{\intercal} \tilde{L}\chi_{\mathcal{S}_i} = \sum_{\substack{u \in \mathcal{S}_i \\ v \in \mathcal{S}_i^c}} \tilde{w}_{uv} = \partial \mathcal{S}_i,
\end{align*}
is the boundary of the set $\mathcal{S}_{i}$ on the graph $\tilde{\mathcal{G}}$, that is, the sum of the edge weights leaving $\mathcal{S}_i$ or the total dynamic coupling between $\mathcal{S}_i$ and $\mathcal{S}_i^{c}$,
\begin{align*}
    \chi_{\mathcal{S}_i}^{\intercal} D \chi_{\mathcal{S}_i} = ||\chi_{\mathcal{S}_i}||_{D}^{2} = \sum\limits_{k = 1}^{n}D_{k}[\chi_{\mathcal{S}_i}]_{k}^{2} = \sum\limits_{k \in \mathcal{S}_{i}}D_{k}
\end{align*}
is the total damping in the subgraph induced by the set $\mathcal{S}_{i}$. Notice when $D_i = 1$ for all $i$, $D = I_n$, and this term is simply the cardinality of the set $\mathcal{S}_{i}$ denoted by $|\mathcal{S}_{i}|$.

Good clusters tend to have small values of $\phi(\cdot)$, therefore, the network clustering problem can be framed as the optimization problem of minimizing the maximum of $k$ measures of badness, this is the $k$-way partitioning problem defined as
\begin{align}
    \rho(k) := \min_{\mathcal{S}_1 ,\ldots, \mathcal{S}_k} \; \max\{ \phi (\mathcal{S}_{i}): i = 1,2,\ldots,k \} \label{eqn:clustering_opt}.
\end{align}
The optimization problem (\ref{eqn:clustering_opt}) is an integer program known to be computationally challenging for large scale networks \cite{lee_multi-way_2014} and the size of the feasible set, $k^n$, grows exponentially with the number of nodes in the network. For any choice of $k$, the solution to (\ref{eqn:clustering_opt}) corresponds to finding $k$ characteristic vectors in $\{1, 0\}^{n}$ with disjoint support \cite{cucuringu_scalable_2016} that minimizes $\rho(k)$.\

From a power network control perspective, the $k$ sets obtained from the solution of (\ref{eqn:clustering_opt}) define $k$ loosely coupled\footnote{Clusters where the sensitivity of the real power flows $p_{\ell}$ to the line angles $\theta_{\ell} = \delta_i - \delta_j$, for $i$ and $j$ at the boundaries are minimized.} zones with balanced frequency responsive components. References \cite{alberto_required_2000, sajadi_synchronization_2022} emphasized the importance of frequency responsiveness (network damping) for improving the transient stability of low inertia grids. Clusters defined in this way ensure that disturbances originating in one cluster are unlikely to propagate through the network, since the line flows at the boundaries are relatively less sensitive. Moreover, having balanced frequency-responsive components across clusters also ensures that each cluster is able to respond to local disturbances.

We can obtain an approximate solution to (\ref{eqn:clustering_opt}) by relaxing the binary and disjointedness constraints on the optimization variables and instead find $k$ non-zero vectors $x_{1}, \ldots, x_{k} \in \mathbb{R}^{n}$ that are $D$-orthogonal (i.e., for all $i \neq j$, $x_{i}^{\intercal}D x_{j} = 0$). The relaxed measure of badness for each cluster when $x_i \in \mathbb{R}^n - \{0\}$ is then given by the generalized Rayleigh-Ritz quotient
\begin{align}
    R\left(\tilde{L}, D; x_{i}\right) = \frac{x_{i}^{\intercal} \tilde{L} x_{i}}{x_{i}^{\intercal} D x_{i}}.
    \label{eqn: rayleigh_quotient}
\end{align}
For any $D\succ0$ and non-zero vector $x_i$ , (\ref{eqn: rayleigh_quotient}) is well defined; and the relaxation of the integer program (\ref{eqn:clustering_opt}) is given by
\begin{align}
    \psi(k) := \min_{x_1, \ldots, x_k \in \mathbb{R}^n - \{0\}} \max \left\{ R\left(\tilde{L}, D; x_i\right) : i = 1, 2, \ldots, k \right\} \label{eqn:spectral_relaxation}.
\end{align}
The $k$ vectors minimizing (\ref{eqn:spectral_relaxation}) are precisely the eigenvectors corresponding to the $k$ smallest eigenvalues of the pair $(\tilde{L}, D)$ \cite{cucuringu_scalable_2016} with an optimal value of $\psi^*(k) = \lambda_{k}(\tilde{L}, D)$ -- this is a consequence of the generalized Courant-Fischer minimax 
theorem for the matrix pair $(\tilde{L}, D)$ when $\tilde{L} \succeq 0$ and $D \succ 0$ \cite{avron_generalized_2008, wolkowicz_handbook_2000}. From (\ref{eqn: rayleigh_quotient}), we see that the eigenvalues of the pair $(L, D)$ defined as the ratio of the quadratic forms of symmetric psd matrices, are non-negative, therefore the eigenvalues of the system dynamics matrix satisfy $\lambda_i(J) \leq 0$ for all $i$.

For a given value of $k$, a power network with a small value of $\psi^*(k)$ -- the $k$th eigenvalue of the linearized dynamics -- implies that the network admits to $k$ loosely coupled partitions with balanced total damping. With $\psi^*(k) = 0$ if and only if the algebraic multiplicity of the $\lambda_i(\tilde{L}) = 0$ is $k$, that is, if the network has at least $k$ islands.

\subsection{Generalized Spectral Embedding}
\label{subsec: generalized_spectral_embedding}
Graph embedding is a process of mapping the nodes of a graph to a low dimensional geometric space that preserves the properties of the network. This geometric space, which is generally Euclidean, is such that the distances between coordinates encode the relationships between the nodes in the network. In this section, we use the generalized eigenvectors of the pair $(\tilde{L}, D)$, which encodes properties of the linearized dynamics introduced in Section~\ref{subsec: optimization_problem}, to construct such an embedding. We then describe a method for obtaining $k$ dynamically coherent clusters (zones) of a power network based on this graph embedding.

Given a pair of symmetric matrices $\tilde{L}, D \in \mathbb{R}^{n \times n}$ the generalized eigenvalue problem is the problem of finding nontrivial solutions to the equation
\begin{align*}
    \tilde{L}x_{i} = \lambda_{i} D x_{i}, \quad i \in \{1, \ldots, n\}
\end{align*}
where $\lambda_{i} = \lambda_i(\tilde{L}, D) \in \mathbb{R}$ and $x_{i} \in \mathbb{R}^{n}$ are the $i$th eigenvalue and right eigenvector, respectively. The generalized eigenvalue problem can be thought of as an ordinary eigenvalue problem over a vector space equipped with a $D$-inner product.

Let $X \in \mathbb{R}^{n \times k}$ denote the matrix whose columns are the first $k$ eigenvectors of the pair $(\tilde{L}, D)$, ordered by increasing eigenvalues. \cite{lee_multi-way_2014} showed that normalizing the columns of the matrix $X$ using the $D$-norm concentrates the rows of $X$ in $k$ different directions corresponding to $k$ clusters. Let $U$ denote this column-wise normalized matrix, the rows of $U$ denoted as $U_{i,:}$ are the coordinates of the corresponding nodes of the graph in $\mathbb{R}^{k}$, with the coordinates of nodes belonging to the same cluster roughly pointing in the same direction.

Since the directions of the rows of $U$ encode cluster membership, we can spatially concentrate the $n$ points in $\mathbb{R}^k$ to improve the performance of geometric partitioning algorithms by radially projecting the coordinates represented by each row onto a $(k-1)$-dimensional unit sphere $\mathbb{S}^{k-1}$ centered at the origin. This is achieved by normalizing $U$ row-wise using the $2$-norm. Each row of this matrix represents a node in the graph, and this $k$-dimensional embedding of the nodes (constructed using the generalized eigenvectors of matrices associated with the graph) is known as the \textit{generalized spectral embedding}. Running $k$-means on this embedding recovers the $k$ clusters. Algorithm \ref{alg:clustering-algo} below is a formal description of the generalized spectral clustering for power networks.
\begin{algorithm}
\caption{Generalized Spectral Clustering $(\text{Network Data}, k)$} 
\label{alg:clustering-algo}
\begin{algorithmic}[1] 
\State \textbf{input}: Power system data, number of control zones $k \geq 2$
\State Run basecase ACOPF to obtain $|V^*|, \delta^{*}, \text{ and }\omega$
\State Construct Dynamic Graph $(\tilde{G})$
\State Compute $X \in \mathbb{R}^{n \times k}$, the first $k$ eigenvectors of the pair $(\tilde{L}, D)$
\For{$i = 1, \ldots, k$}
    \State $d_{i} = ||X_{:,i}||_{D}$
    \State $U_{:,i} = X_{:,i}/d_{i}$
\EndFor
\For{$j = 1, \ldots, n$}
    \State $s_{j} = ||U_{j,:}||_{2}$
    \State $U_{j,:} = U_{j,:}/s_{j}$
\EndFor
\State Cluster the points $\left\{ U_{j,:} \right\}_{j = 1}^{n}$ into $k$ clusters using $k$-means
\State Check that cluster defined by the sets $\mathcal{S}_1, \ldots, \mathcal{S}_k$ are connected on $\mathcal{G}$
\State \textbf{return} $\mathcal{S}_1, \ldots, \mathcal{S}_k$
\end{algorithmic}
\end{algorithm}

The number of clusters (the value of $k$) can be determined by the requirements of the underlying control scheme or pre-specified by the system operator. Ideally, $k$ should be such that the first $k$ generalized eigenvalues $\lambda_{1}(\tilde{L}, D), \ldots, \lambda_{k}(\tilde{L}, D)$ are small while $\lambda_{k+1}(\tilde{L}, D)$ is relatively large. In the absence of a prespecified value of $k$, a common heuristic for choosing $k$ is to select $k$ with the largest relative spectral gap, that is
\begin{align*}
    k = \arg \max_{k \geq 2}\left(\frac{\lambda_{k+1}(\tilde{L}, D) -\lambda_{k}(\tilde{L}, D)}{\lambda_{k}(\tilde{L}, D)}\right).
\end{align*}
There are several justifications for this choice of $k$. First, a large relative spectral gap ensures that the community structure of the network (the number of clusters) is robust across a wide range of steady-state operating points. Appendix~\ref{appendix: eigenvalue_perturbation} and \ref{appendix: eigenspace_perturbation} provides an explicit bound on the worst-case change in the eigenvalues of the linearized oscillator model in terms of the spectral radius of the perturbation to the Laplacian of the dynamic graph and a measure of the definiteness of the pair $(\tilde{L}, D)$. Secondly, choice of $k$ ensures that the worst of the $k$ clusters( as evaluated using $\phi(\cdot)$) returned by the method is good. We note that the damping coefficients of the oscillators are identical (i.e. $D = \alpha I$, for some $\alpha>0$) the relative spectral gap does not depend on the damping coefficient $\alpha$.

The embedding of the network constructed using Algorithm~\ref{alg:clustering-algo} is unique when the $k$ smallest eigenvalues of the matrix pair $(\tilde{L}, D)$ are distinct, since the eigenspaces corresponding to distinct eigenvalues are disjoint subspaces, the $k$ eigenvectors computed at step~4 of Algorithm~\ref{alg:clustering-algo} are therefore unique up to scaling.
We note that the dynamic graph $\tilde{\mathcal{G}}$ can be constructed solely from the network parameters and measurements without the need to solve an AC optimal power flow problem. This can significantly reduce the computational complexity of the method to that of computing the eigenvectors of the generalized eigenvalue problem -- the most computationally expensive step -- which can be computed exactly in polynomial time $\mathcal{O}(n^3)$ using QZ factorization, or approximately using the inverse power method which converges to a sufficiently good solution in $\mathcal{O}(\log(n))$\cite{cucuringu_scalable_2016}.


\section{Results}
\label{sec: results}
In this section, we evaluate the performance of the generalized spectral clustering approach proposed in Section~\ref{sec: clustering} on the IEEE 30-bus test system.
The network parameters, basecase system load, bus voltage magnitudes, and angles are obtained from MATPOWER's \textit{case$30$} \cite{zimmerman_matpower_2011}. The network has $|\mathcal{V}_G \cup \mathcal{V}_R| = 6$ generator buses, $|\mathcal{V}_L| = 24$ load buses, and $|\mathcal{E}| = 41$ transmission lines. The edge weights $\tilde{w}_{ij}$ of the branches of the dynamic graph $\tilde{\mathcal{G}}$ are obtained from the basecase ACOPF solution (steady-state voltage magnitudes and angles) of the system and line parameters.

For nodes $i \in \mathcal{V}_{G} \cup \mathcal{V}_R$, we choose the inertia coefficients $M_{i} \sim \text{uni}(0.5, 2)$, and $D_{i} \sim \text{uni}(25, 30)$. These damping and inertia values are consistent with those in \cite{dorfler_synchronization_2012}, where, for synchronization problems, the impact of the synchronous generator's excitation control are accounted for in the damping 
term. For load buses $i \in \mathcal{V}_{L}$, $D_{i} \sim \text{uni}(1.0, 1.5)$, reflecting the fact that loads are typically less frequency responsive compared to synchronous generators with damper windings and excitation control and inverters implementing a frequency-droop control law.
\begin{figure}[!th]
    \centering
    \begin{subfigure}[b]{0.49\linewidth}
        \centering
        \includegraphics[width=\linewidth]{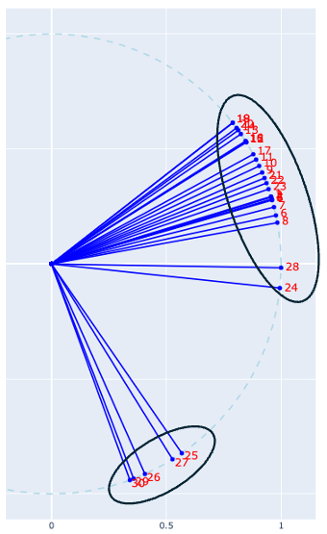}
        \caption{}
        \label{fig: 2D_spectral_embedding}
    \end{subfigure}
    \hfill
    \begin{subfigure}[b]{0.49\linewidth}
        \centering
        \includegraphics[width=\linewidth]{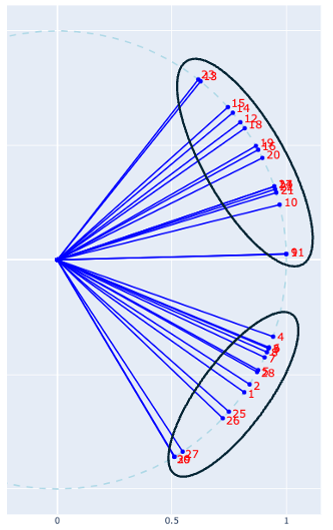}
        \caption{}
        \label{fig: 2D_generalized_spectral_embedding}
    \end{subfigure}
    \caption{$2$-D Embedding of the IEEE 30-bus test network using the (a) eigenvectors of $\tilde{L}$, (b) the eigenvectors of the pair $(\tilde{L}, D)$. 
    The total edge weights cut by the spectral and generalized spectral clustering solutions on $\tilde{\mathcal{G}}$, are 5.04 and 12.55, respectively, and the corresponding total damping in each cluster is 157.89 and 30.01 for spectral clustering, and 91.33 and 96.57 for generalized spectral clustering.
    }
    \label{fig: 2D_embeddings}
\end{figure}

We first consider the clustering solution for $k = 2$, this choice of $k$ allows us to visualize the low-dimensional embedding of the network. Fig.~\ref{fig: 2D_spectral_embedding} shows the embedding obtained using the eigenvectors of the Laplacian matrix $\tilde{L}$, which only considers the connectivity of the graph. In contrast, Fig.~\ref{fig: 2D_generalized_spectral_embedding} is the embedding of the network obtained from the eigenvectors of the matrix pair $(\tilde{L}, D)$, which incorporates both the effect of the connectivity of the network and the impact of the heterogeneity of the damping coefficients of the oscillators. We see in Fig. \ref{fig: 2D_embeddings} that the time constants $D_i$ play a major role in shaping the network embedding. The $2$ clusters obtained from running $k$-means on the respective embeddings have been highlighted in Fig.~\ref{fig: 2D_embeddings}. As expected, the coordinates in the embedding (projected on the unit circle) of nodes belonging to the same cluster are aligned and point roughly in the same direction.

The clusters in Fig.~\ref{fig: 2D_spectral_embedding}, identified using the spectrum of the Laplacian matrix $\tilde{L}$, can differ significantly from those obtained using the spectrum of the linearized coupled oscillator equations, which govern the evolution of voltage angles around a synchronous solution -- and hence branch power flows. Comparing the clusters identified on the graph $\tilde{\mathcal{G}}$ using the respective embeddings Fig.~\ref{fig: 2D_embeddings}, we observe that the clusters on the graph obtained using the eigenvectors of $(\tilde{L}, D)$ cuts slightly more edges in an attempt to ensure that the total edge weights on the boundary of the clusters $\tilde{\mathcal{G}}_{\mathcal{S}}$ and $\tilde{\mathcal{G}}_{\mathcal{S}^c}$ are minimized while ensuring that the total frequency responsive components in the resulting clusters are balanced.
\begin{figure}[htbp]
    \centering
    \includegraphics[width=0.8\linewidth]{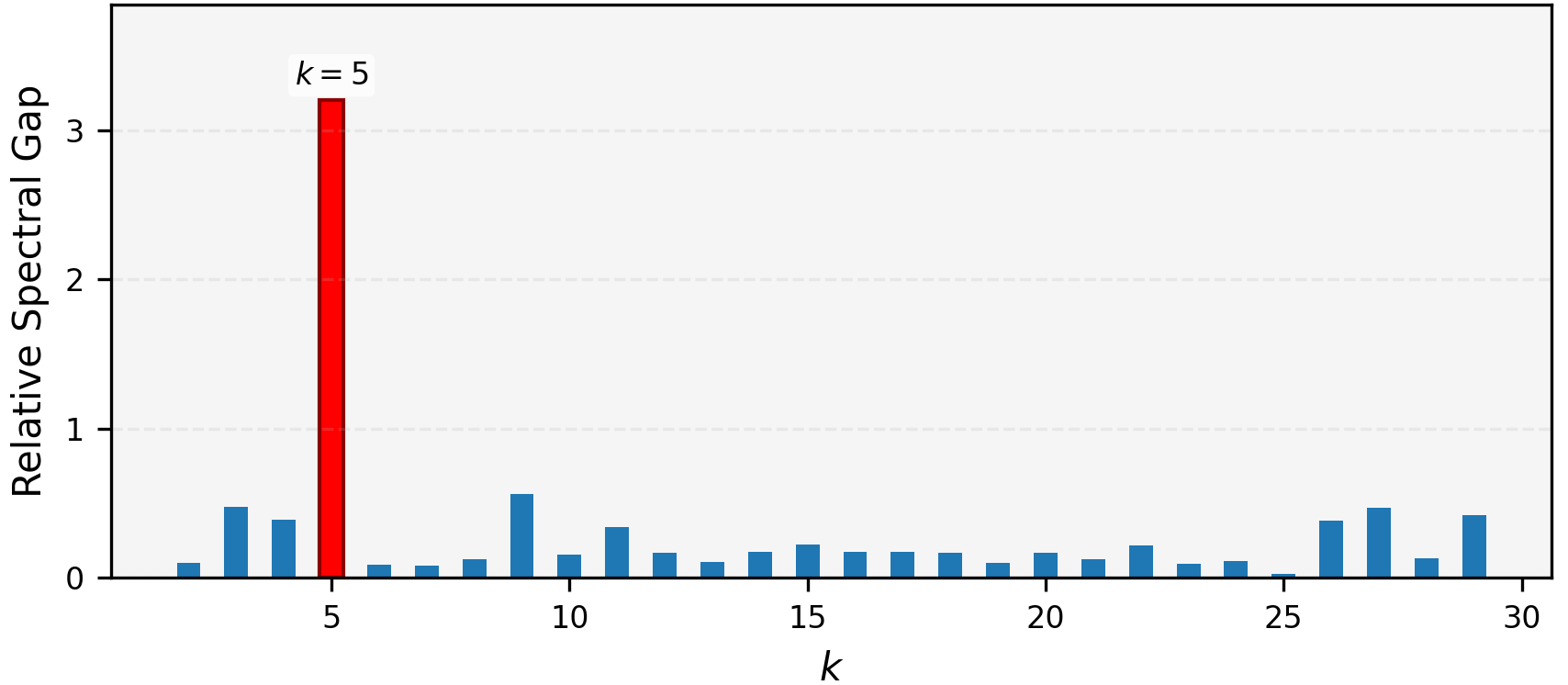}
    \caption{Relative spectral gap of $\tilde{L}x = \lambda D x$.}
    \label{fig: relative_spectral_gap}
\end{figure}

To address the question of a good choice of $k$ for this network, we plot the relative spectral gap of the eigenvalues of $(\tilde{L}, D)$ in Fig.~\ref{fig: relative_spectral_gap}. We see that $k=5$ has the largest relative spectral gap, suggesting that there are five dynamically coherent groups of nodes in the network.
The resulting clusters from Algorithm~\ref{alg:clustering-algo} for $k=5$ are shown on the dynamic graph in Fig. \ref{fig: dynamic_graph} below. The thickness of each edge on the figure reflects the strength of coupling between the pair of connected nodes, while the node sizes have been scaled to approximately reflect the magnitude of the effective damping $D_i$ at each node (which can be thought of as node weights in the linearized model).
\begin{figure}[htbp]
    \centering
    \includegraphics[width=1.0\linewidth]{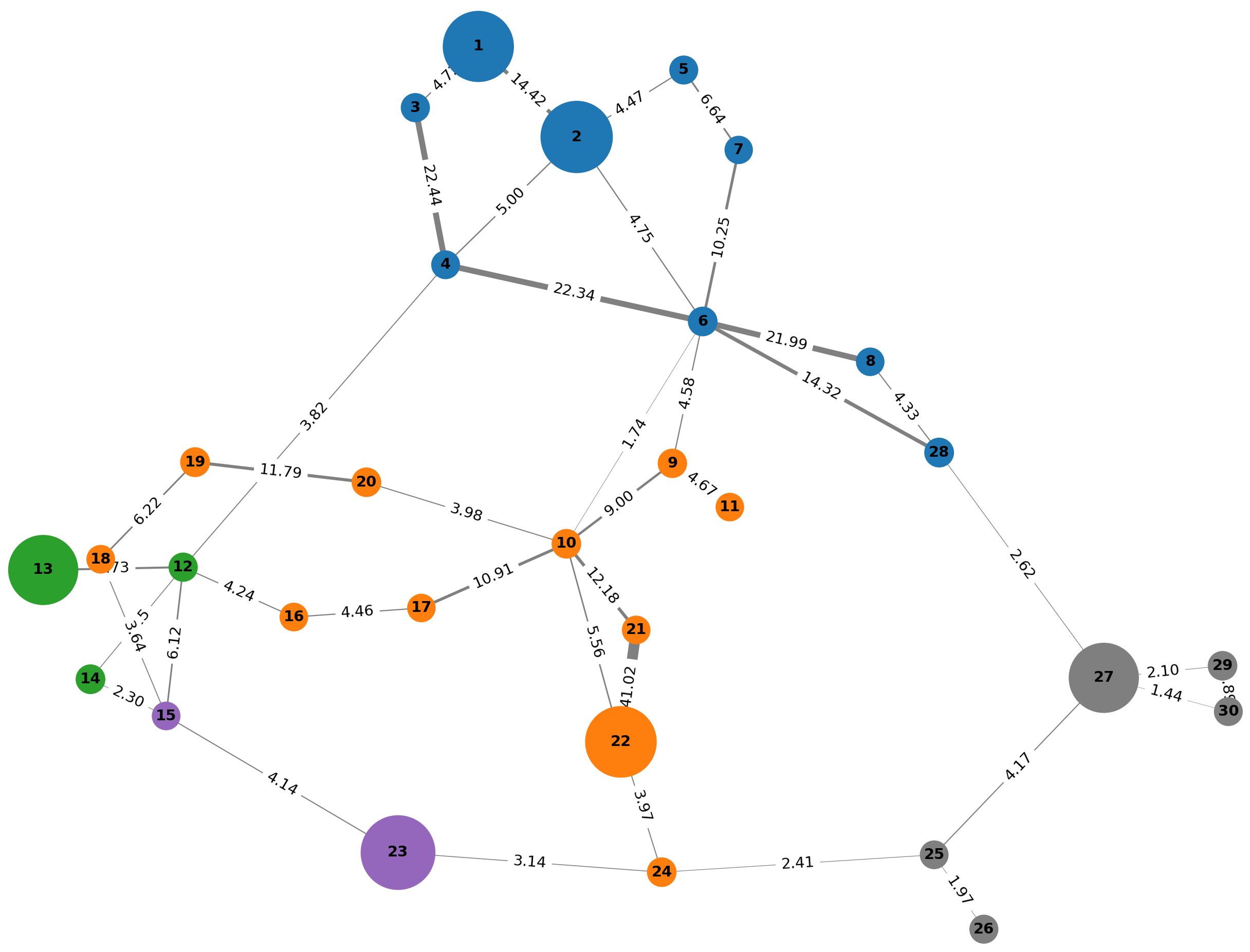}
    \caption{The dynamic graph $\tilde{G}$ of the IEEE \textit{case}$30$ showing five clusters. Note that the position of the nodes on the graph do not reflect the physical proximity of the buses in the network.}
    \label{fig: dynamic_graph}
\end{figure}
Table \ref{tab: clustering_solution} shows the boundaries of and the total damping within each cluster. The quality of each of the resulting clusters, determined from the value of $\phi(\cdot)$, is shown in Table \ref{tab: clustering_solution}.
\begin{table}[h]
\centering
\caption{Five Control Zones (Clusters) on the IEEE 30-bus System}
\label{tab: clustering_solution}
\resizebox{\linewidth}{!}{
    \begin{tabular}{l|c|c|c|c}
    \hline
     \textbf{Clusters} & \textbf{Nodes} & \textbf{Boundary} & \textbf{Total Damping} & \textbf{$\phi(\mathcal{S}_{i})$} \\
    \hline
    \text{Blue ($\mathcal{S}_{1})$}    & 1 -- 8, 28   & 12.76    & 61.32 & 0.21    \\
    \hline
    \text{Orange ($\mathcal{S}_{2})$}   & 9 -- 11, 16 -- 22, 24   & 19.76    & 38.72 & 0.51    \\
    \hline
    \text{Green ($\mathcal{S}_{3})$}   & 12 -- 14   & 16.48    & 27.76 & 0.59    \\
    \hline
    \text{Gray ($\mathcal{S}_{4})$}   & 25 -- 27, 29, 30   & 5.04    & 30.01 & 0.17    \\
    \hline
    \text{Purple ($\mathcal{S}_{5})$}   & 15, 23   & 15.20    & 30.08 & 0.51    \\
    \hline
    \end{tabular}
}
\end{table}
We see that the best clusters $\mathcal{S}_4$ and $\mathcal{S}_1$ with $\phi(\cdot)$ values of $0.17$ and $0.21$, respectively, are 
clearly distinguishable
from the layout of the graph in Fig. \ref{fig: dynamic_graph}. Cluster defined by $\mathcal{S}_3$ has the highest measure of badness which captures the combination of relatively low total damping and strong coupling to adjacent clusters $\mathcal{S}_2$ and $\mathcal{S}_5$, with total couplings of $4.24$ and $8.42$, respectively.
We note that each of the resulting clusters in Fig.~\ref{fig: dynamic_graph} have at least one generator node, with more generators improving 
the measure of goodness of the resulting cluster as is the case for $\mathcal{S}_1$.

\begin{figure}[htbp]
    \centering
    \includegraphics[width=0.5\textwidth]{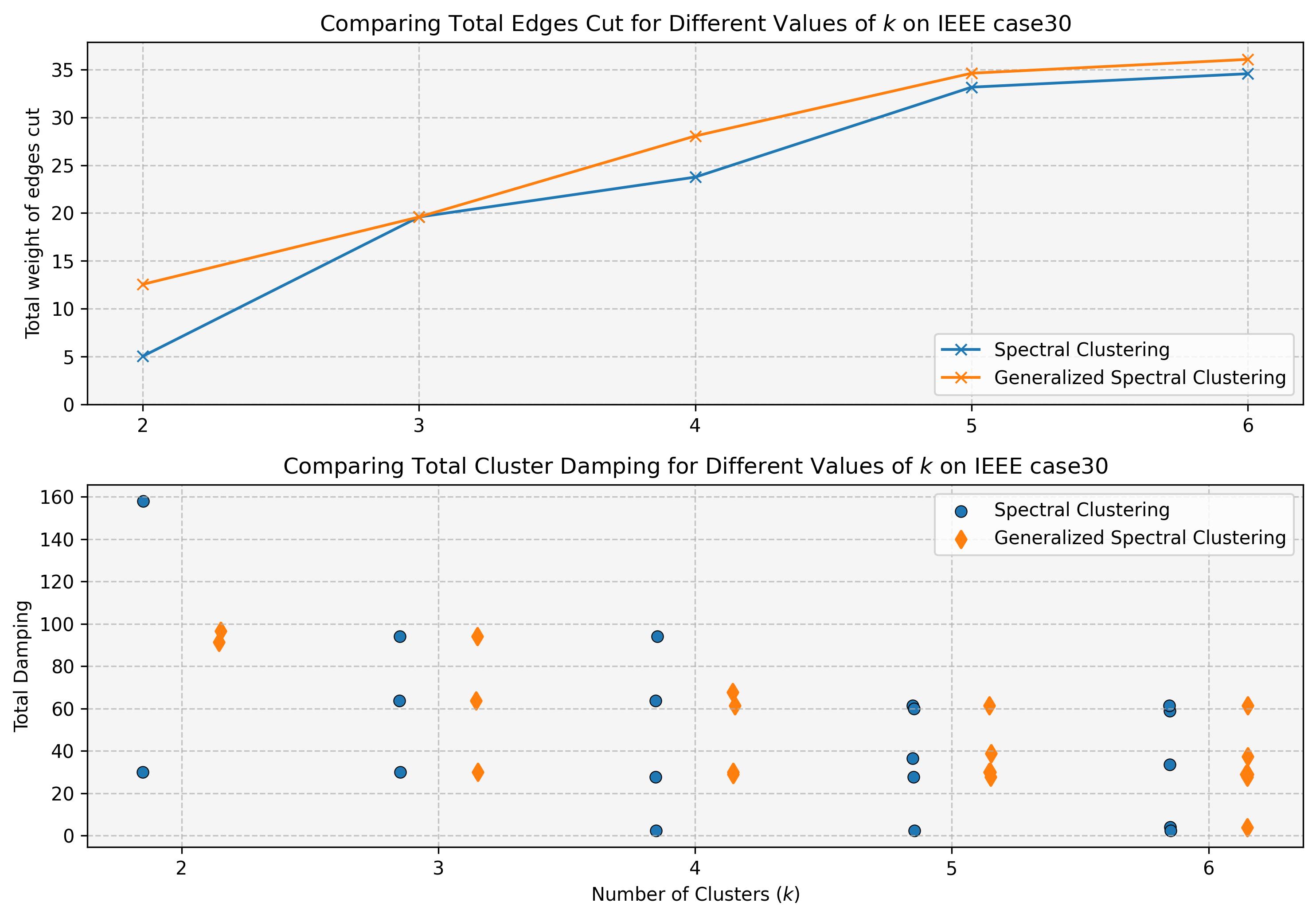}
    \caption{Total edges cut and total cluster damping for different values of $k$.}
    \label{fig: cuts_and_damings}
\end{figure}

While the network embeddings are challenging to visualize for $k>3$, Fig. \ref{fig: cuts_and_damings} compares the total edges cut and the total damping in the clusters obtained using the eigenvectors of $\tilde{L}$ versus the eigenvectors of $(\tilde{L},D)$ for values of $k$ ranging from $2$ to $6$. The values of $k$ considered in this analysis are limited to $k \leq 6$, as larger values of $k$ tend to return clusters with multiple connected components and/or singletons. Fig. \ref{fig: cuts_and_damings} shows that, consistently, when the solutions differ, the partitions obtained via the eigenvectors of the generalized eigen problem tends to cut slightly more edges to ensure that the overall distribution of damping across the resulting clusters is balanced. Specifically, for $k>3$ one or more clusters returned using the eigenvectors of $\tilde{L}$ have no generators, while each cluster obtained using the generalized embedding balances damping across clusters by including at least one generator node for $k$ up to $5$. We observe that for $k=3$, the two embeddings return the same solution. We note that when the time constants across the network are uniform, the two embeddings are identical, and $k$-means returns the same clustering solution for all $k$.

\begin{figure}[htbp]
    \centering
    \includegraphics[width=0.9\linewidth]{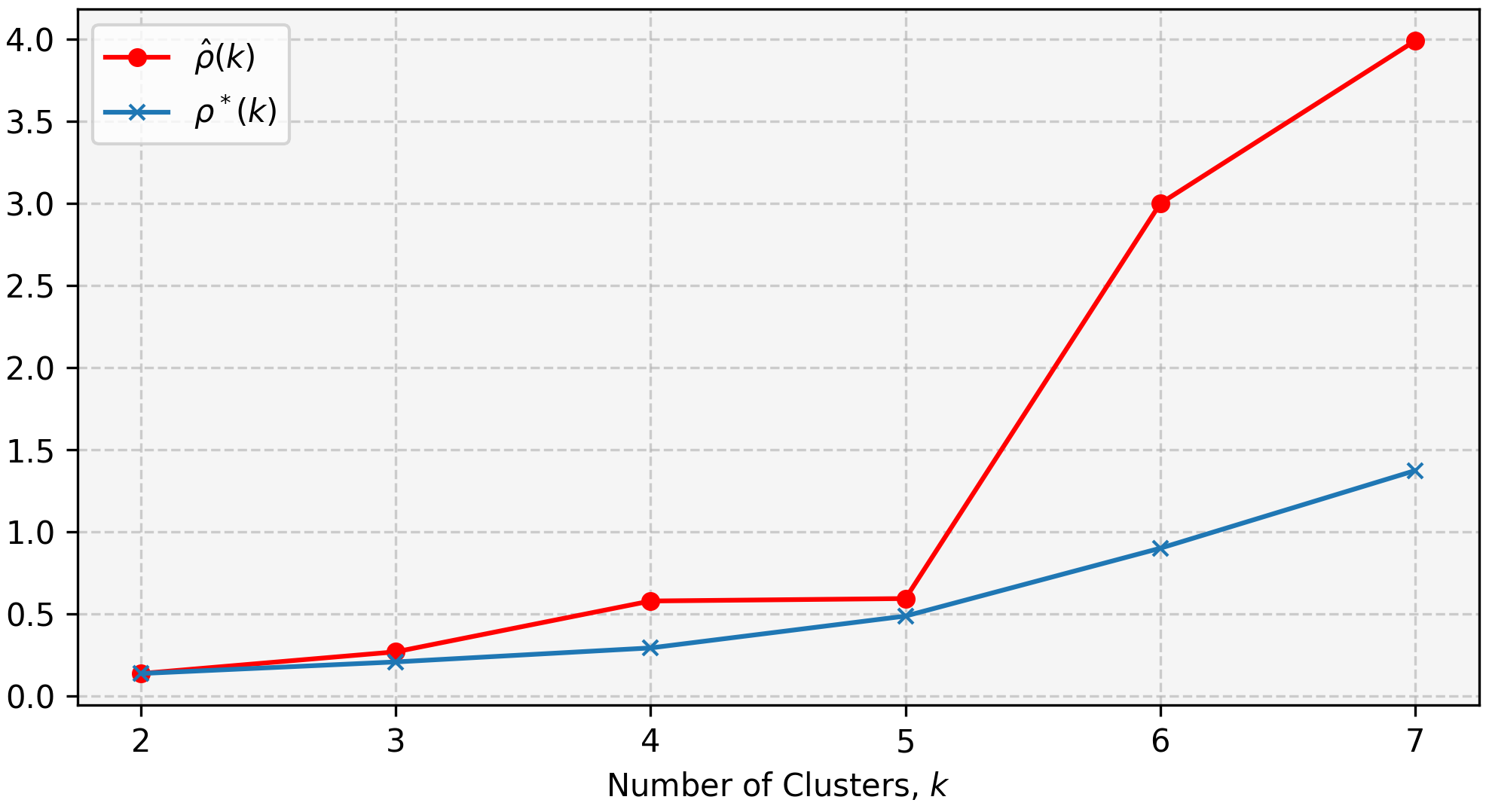}
    \caption{The optimal value $\rho^*(k)$ of (\ref{eqn:clustering_opt}) and objective value $\hat{\rho}(k)$ of the generalized spectral clustering solution for different values of $k$.}
    \label{fig: optimal_rho_ks}
\end{figure}

We also compare the performance of the proposed method to the globally optimal solution of the integer problem in ~\ref{eqn:clustering_opt}. We denote the optimal value of (\ref{eqn:clustering_opt}) for a given value of $k$ as $\rho^*(k)$ and the objective value of (\ref{eqn:clustering_opt}) obtained using the eigenvectors of $(\tilde{L}, D)$
-- the maximum of $k$ measures of \textit{badness} of the resulting clusters -- as $\hat{\rho}(k)$. From Fig. \ref{fig: optimal_rho_ks} we see that for $k \leq 5$, the objective values of the solutions obtained using Algorithm \ref{alg:clustering-algo} are close to the globally optimal values, with the approximate solution coinciding with the true minimizer of the integer program for $k=2$.

Finally, for larger values of $k$, the objective values obtained from clustering using the eigenvectors of $(\tilde{L}, D)$ (shown in red) deviate significantly from the global optima (shown in blue) for this network. The relatively steep change in the objective values $\hat{\rho}(k)$ for $k \geq 6$ also confirms that the network does not admit to $k\geq 6$ good clusters.

\section{Numerical Validation}
\label{sec: numerical_validation}
\subsection{Network Dynamics Simulations}
We validate the dynamic coherence of the clusters identified in section~\ref{sec: results} for $k=5$ by numerically simulating the dynamics of the generalized coupled oscillator model (equations \ref{eq: 1} - \ref{eq: 2}) in response to random disturbance in the natural frequency (net power injection) $\omega_i$ of the $i$th oscillator. 
These disturbances can represent network transients caused by deviations from scheduled net injections due to fluctuations (stochasticity) in renewable generation or large load changes in the network. The simulations use the inertia and damping parameters specified in Section~\ref{sec: results}.

For each $i\in \mathcal{V}$, we perturb the net power injection $\omega_i$ and determine the effect of this disturbance on the network by measuring the pair-wise coherence between the angular frequency trajectories $\dot{\delta}_i(t)$ and $\dot{\delta}_j(t)$ for all $j$ in the synchronously rotating reference frame of frequency $\omega_{\text{sync}}$ defined as
\begin{align*}
    \cos(\theta_{ij}) = \frac{\langle e_i(t) | e_j(t)\rangle}{||e_{i}(t)||_2 \cdot ||e_j(t)||_2},
\end{align*}
where $e_i (t) = \dot{\delta}_i (t) - \omega_{\text{sync}}$, and $\omega_{\text{sync}} = (\sum_{i} \omega_i)/(\sum_i D_i)$.

\begin{figure}[htbp]
    \centering
    \includegraphics[width=1.0\linewidth]{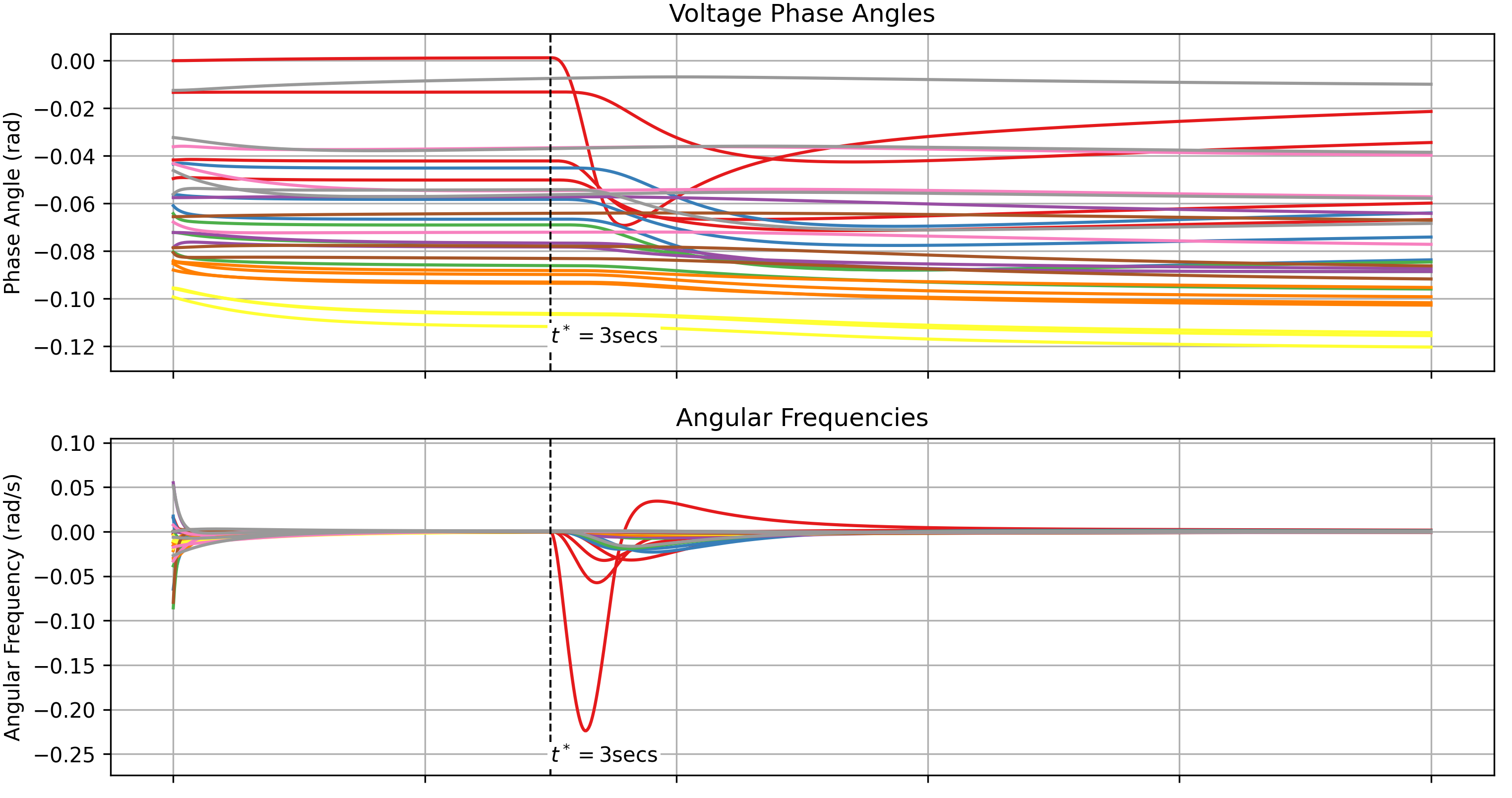}
    \caption{Phase and frequency trajectories following a disturbance at node $1$.}
    \label{fig: phase_and_frequency_traj}
\end{figure}

We initialize the simulation at $t = 0$ sec using the voltage phase angles $\delta(0) \in \mathbb{T}^n$ obtained from the ACOPF solution of the basecase, and angular frequencies $\dot{\delta}(0) = \mathbf{0}$. At time $t^* = 3$ sec in the simulation interval, following an initial synchronization, a random continuous-time disturbance $\xi(t)$ is added to $\omega_i$ for a duration of $0.5$ sec. Fig. \ref{fig: phase_and_frequency_traj} shows the impact of such a disturbance at node $1$ on the voltage phase angle $\delta(t)$ and angular frequency $\dot{\delta}(t)$ trajectories of all the oscillators in the network.

\begin{figure}[htbp]
    \centering
    \includegraphics[width=1.0\linewidth]{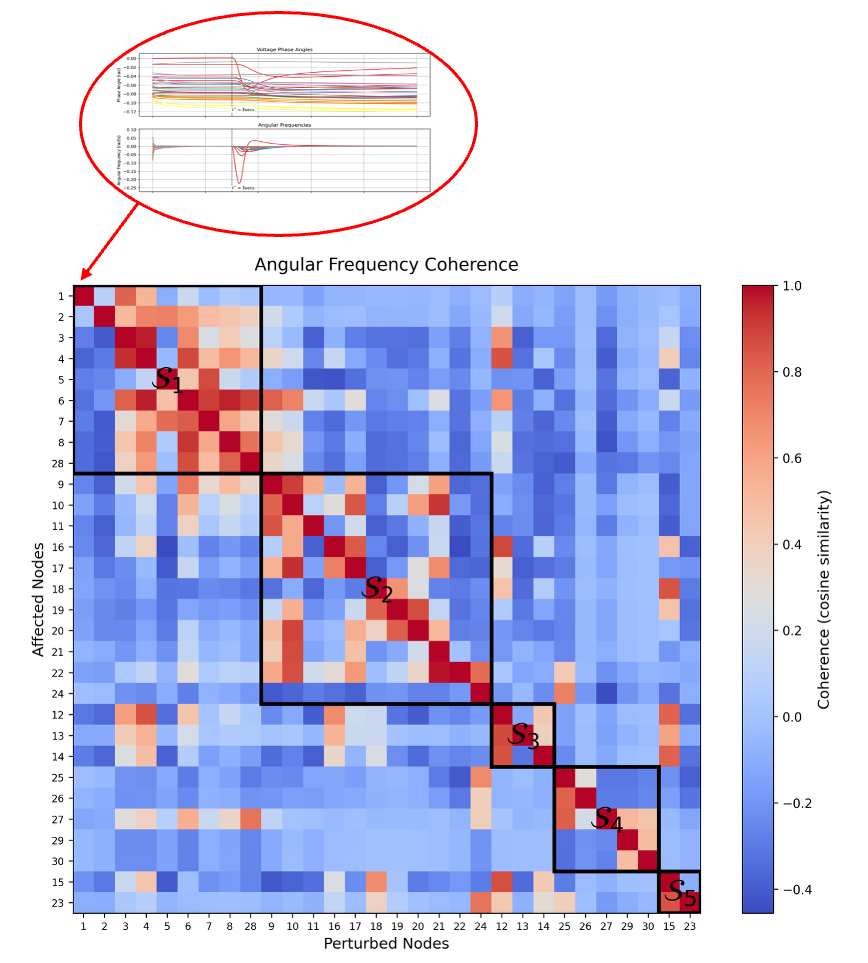}
    \caption{Angular frequency coherence heatmap of the IEEE 30-bus network.}
    \label{fig: coherence_heatmap}
\end{figure}

The overall impact of such perturbations across the network is shown on the coherence heatmap Fig. \ref{fig: coherence_heatmap}, where the $ij$-th entry reflects the coherence between $\dot{\delta}_i(t)$ and $\dot{\delta}_j(t)$ following a disturbance at node $i$. The rows and columns of the heatmap have been permuted according to cluster membership, so that nodes belonging to the same cluster are adjacent (and they have been labeled accordingly).
Notice that the coherence heatmap is not symmetric, since the effect of a disturbance at node $i$ on node $j$ depends not only on the paths, through the network, between the source and sink node, but also on the damping in the neighborhood of the source node which determines how quickly disturbances are attenuated before it spreads through the network.

As expected, the coherence matrix is block-diagonally dominant, highlighting the modular structure of the network. The voltage angle trajectories of nodes within each identified cluster exhibit high intra-cluster coherence, in contrast to those outside the cluster. The block-diagonal structure also confirms that disturbances originating within each of the identified cluster is mostly localized to that cluster. The weaker the inter-cluster coupling (i.e., the sensitivity of real power flow to the line angle), the closer the coherence matrix is to being block-diagonal; in the limiting case of a network with $ks$ islands, the matrix is entirely block-diagonal.

\subsection{Robustness of Clustering Solution}
To determine the robustness of the clusters identified in Section~\ref{sec: results} using the spectrum of the linearized coupled oscillator dynamics equation, we consider $1,000$ random steady-state operating conditions of the test network.

For each load bus in the network, we generate $1,000$ realizations of uniformly distributed load demands as follows
\begin{align*}
    \omega_i[k] = \omega_i^0 + \zeta_i[k] \qquad \forall i\in \mathcal{V}_L
\end{align*}
where $\omega_i^0$ is the nominal demand (in MW) at node $i$, $\zeta_i \sim \mathcal{N}(0, \sigma^2)$ with variance $\sigma^2 = 5$ MW. We then compute the corresponding net-injections of the generator buses, and the voltage magnitudes and angles across the network by solving the ACOPF problem.
Randomizing the load demands induces a structured randomization in the realized steady states of the network and the corresponding Laplacian, ensuring the operational feasibility of the resulting states. 
\begin{figure}[htbp]
    \centering
    \includegraphics[width=1.0\linewidth]{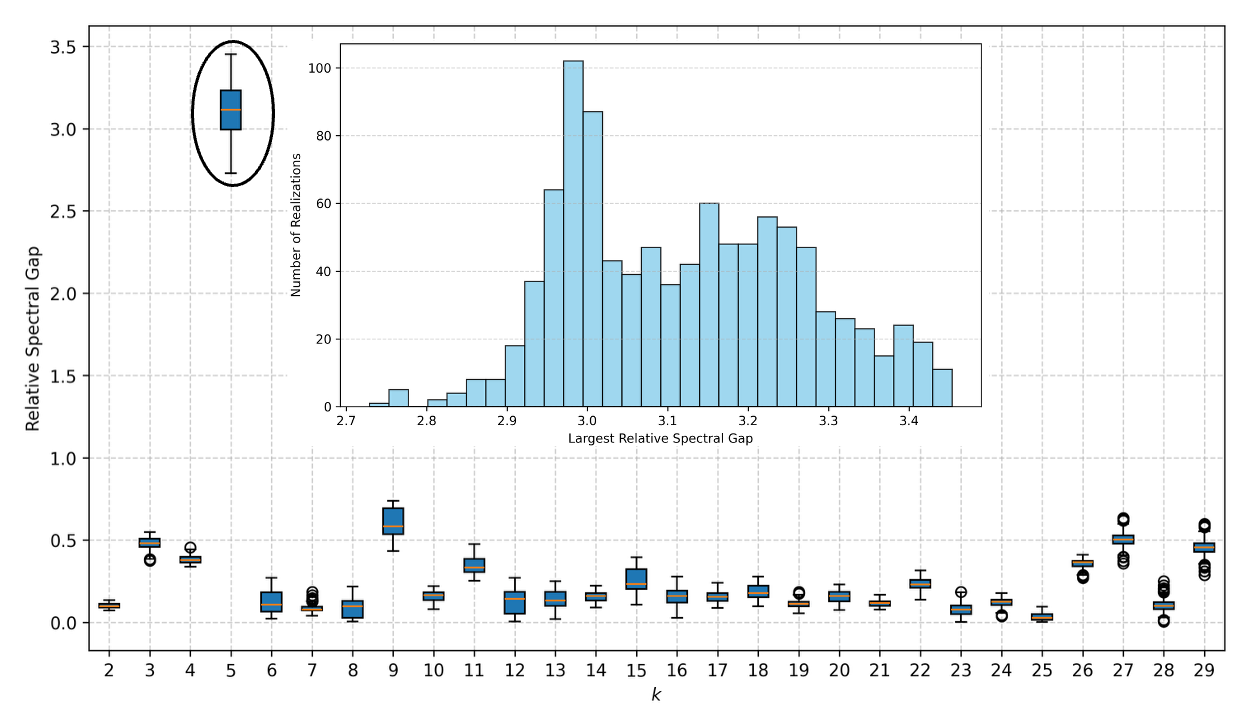}
    \caption{Relative spectral gaps for $1,000$ random steady-states of the IEEE 30-bus network.}
    \label{fig: randomized_spectral_gap}
\end{figure}
Across all considered operating points, the largest relative spectral gap was consistently between the $5$th and $6$th generalized eigenvalues. Fig.~\ref{fig: randomized_spectral_gap} shows the relative spectral gaps across the realized states. The distribution of the largest relative spectral gap for the network has a mean of $3.1218$ and variance of $0.0213$ (relative variance of $0.007$), confirming that choice of $k=5$ clusters for this network is robust to variability in system operating conditions.

The nodal assignments to individual clusters are also relatively stable. Specifically, the nodal assignments identified in Section~\ref{sec: results} are consistent under randomized changes to the operating point with the following exceptions:
\begin{enumerate}
    \item Node $15$ which is assigned to $\mathcal{S}_5$ (purple) in $99.5\%$ of the realizations and to $\mathcal{S}_3$ (green) in only $0.5\%$, and
    \item Nodes $18$ and $24$ belonging to clusters $\mathcal{S}_2$ (orange) and $\mathcal{S}_5$ (purple) $97.90\%$ and $2.1\%$ of the realizations, respectively.
\end{enumerate}
The robustness of the resulting clusters ensures that the identified clusters remain stable under changes in the network's steady-state operating points, precluding the need to re-solve the clustering problem for slight changes in the network's loading condition. This reduces the need to periodically switch the communication/control structure of distributed control schemes like coalition MPC \cite{chanfreut_fast_2022}.

\section{Conclusion}
\label{sec: conclusion}
In this work, we presented an approach based on the framework of coupled oscillators for identifying dynamically coherent control zones in low-inertia power networks. We established the connection between the spectrum of the linearized voltage angle dynamics matrix with heterogeneous time constants and the broader framework of generalized spectral clustering, and showed that an embedding of the network using the eigenvectors of the linearized system dynamics matrix results in a certain decomposition of the network.

We demonstrate our method on the IEEE 30-bus network and compare the clusters obtained from the eigenvectors of the Laplacian matrix of the dynamic graph to those obtained using the eigenvectors of the linearized system dynamics matrix. The results show that clustering solutions using the eigenvectors of the dynamics matrix (i.e., the eigenvectors of the generalized eigenvalue problem) attempt to balance a trade-off between cutting branches where the sensitivity of real power flow to line angle is small and ensuring that the resulting clusters have balanced frequency-responsive components. The secondary objective of balancing the amount of frequency-responsive components ensures that the clusters identified by this method tend to include a generator bus, a factor that contributes to the operational resilience of each cluster. Together, these objectives ensure that disturbances originating within each cluster are reasonably contained.

We validate the dynamic performance of the resulting clustering solution by simulating the impact of random perturbations in the natural frequencies (net power injections) on the frequency dynamics of the network. Results showed that the impact of disturbances originating within each of the identified clusters is indeed localized to that cluster, and that the frequency dynamics of nodes belonging to the same cluster are coherent. This property allows the resulting control zones to serve as an effective design heuristic for structuring the feedback matrix in distributed controller synthesis problems. In addition, the proposed clustering method could facilitate model reduction by allowing groups of dynamically coherent nodes to be represented as a single node in a reduced network model for studying the aggregated dynamics of the network. Results show that the coherence structure -- the number of clusters and the individual node assignments to each cluster -- is robust across a wide range of steady-state system operating conditions.


\appendices
\section{A Perturbation Bound on Eigenvalues}
\label{appendix: eigenvalue_perturbation}
Given a symmetric pair $(L, D)$, let $\sigma(L, D) = \{\lambda_1, \lambda_2, \ldots, \lambda_n\}$ denote the set of the generalized eigenvalues. The pair $(L, D)$ is \textit{definite} if
\begin{align*}
    \mu(L,D) = \min_{\substack{x \in \mathbb{R}^n\\ ||x||_2 = 1}}\sqrt{(x^{\intercal}Lx)^2 + (x^{\intercal}Dx)^2} > 0
\end{align*}
Consider symmetric perturbations $\Delta L$ and $\Delta D$ to these matrices. Define $\widehat{L} = L + \Delta L$ and $\widehat{D} = D + \Delta D$, and let $\sigma(\widehat{L}, \widehat{D}) = \{\widehat{\lambda}_1, \widehat{\lambda}_2,\ldots, \widehat{\lambda}_n\}$ denote it's eigenvalues. The chordal distance between the non-zero generalized eigenvalues $\lambda$ and $\widehat{\lambda}$ is defined as
\begin{align*}
    \text{dist}_c(\lambda, \widehat{\lambda}) = \frac{|\lambda - \widehat{\lambda}|}{\sqrt{|\lambda|^2 + 1} \sqrt{|\widehat{\lambda}|^2 + 1}}.
\end{align*}
Theorem $5$ in Section $21.4$ of \cite{hogben_handbook_2013}: Suppose $(L, D)$ is a definite pair, if $\widehat{L}$ and $\widehat{D}$ are symmetric and $||[\Delta L, \Delta D]||_2 < \mu(L, D)$, then $(\widehat{L}, \widehat{D})$ is also a definite pair and there exists a permutation $\tau$ of the indices $\{1, 2, \ldots , n\}$ such that
\begin{align*}
    \max_{1\leq j \leq n} \text{dist}_c(\lambda_j, \widehat{\lambda}_{\tau(j)})
    \leq \frac{||[\Delta L, \Delta D]||_2}{\mu(L,D)},
\end{align*}
where $||[\Delta L, \Delta D]||_2$ is the operator $2$-norm of the stacked perturbation matrix. If we assume that the damping coefficients of the oscillators are constant, then linearizing the network around different operating points correspond to symmetric perturbations to the Laplacian matrix of the dynamic graph. It follows that the worst-case change in the eigenvalues of the pair $(L, D)$ (and the eigenvalues of the system dynamics matrix $J$) is bounded as
\begin{align*}
    \max_{1\leq j \leq n} \text{dist}_c(\lambda_j, \widehat{\lambda}_{\tau(j)}) &\leq \frac{||[\Delta L, \mathbf{0}]||_2}{\mu(L,D)}
    = \frac{\rho(\Delta L)}{\mu(L,D)} \leq \frac{\rho(\Delta L)}{\min_i D_{ii}},
\end{align*}
where $\rho(\Delta L) = \max_{i}|\lambda_{i}(\Delta L)|$ is the spectral radius of the perturbation $\Delta L$. Therefore, provided the pair $(L,D)$ is definite, the change in the eigenvalues is bounded by the ratio $\rho(\Delta L)/\min_i D_{ii}$.

\section{Perturbation bound on the Eigenspace}
\label{appendix: eigenspace_perturbation}
Theorem $3.5$ \cite{stewart_matrix_1990}: Let $(L, D)$ be a definite pair. Let the columns of $Z_1$ span an eigenspace of $(L, D)$. Then there is a matrix $Z_2$ such that $[Z_1, Z_2]$ is nonsingular and the pair $(L, D)$ has \textit{spectral resolution}
\begin{align*}
    \begin{bmatrix}
        Z_1^H\\
        Z_2^H
    \end{bmatrix}L[Z_1, Z_2] = \begin{bmatrix}
        L_1 & \mathbf{0}\\ \mathbf{0} & L_2
    \end{bmatrix}
\end{align*}
and
\begin{align*}
    \begin{bmatrix}
        Z_1^H\\
        Z_2^H
    \end{bmatrix}D[Z_1, Z_2]  = \begin{bmatrix}
        D_1 & \mathbf{0}\\ \mathbf{0} & D_2
    \end{bmatrix}
\end{align*}
Moreover, $Z_1$ and $Z_2$ may be chosen so that $L_1, L_2, D_1, D_2$ are diagonal (i.e., the columns of $[Z_1, Z_2]$ are eigenvectors).

Theorem 3.1 \cite{sun_perturbation_1983}: Let $(L, D), (\widehat{L}, \widehat{D})$ be definite matrix pairs, let $Z$ and $\tilde{Z}$ be as defined in Theorem $3.5$. Define
\begin{align*}
    \delta = \min_{i,j}\{\text{dist}_c(\lambda_i, \widehat{\lambda}_j): \; &\lambda_i\in \sigma(L_1, D_1), \; \widehat{\lambda}_j \in \sigma(\widehat{L}_2, \widehat{D}_2)\}.
\end{align*}
If $\delta > 0$, then the difference between the subspaces $\mathcal{R}(Z_1)$ and $\mathcal{R}(\widehat{Z}_1)$
\begin{align}
    \label{eigenspace_bound} ||\sin \Theta_1||_F \leq \frac{||(L, D)||_2}{\mu(L, D) \mu(\widehat{L}, \widehat{D})}\cdot\frac{||(\Delta L Z_1, \Delta D Z_1)||_F}{\delta},
\end{align}
where
\begin{align*}
    ||(L, D)||_2 &= \sqrt{||L^2 + D^2||_2},\\
    ||(\Delta L Z_1, \Delta D Z_1)||_F &= \sqrt{||\Delta L Z_1||_F^2 + ||\Delta D Z_1||_F^2}.
\end{align*}
Given an $\ell$-dimensional eigenspace $\mathcal{R}(Z_1)$ of the pair $(L, D)$ if the corresponding eigenvalues of $(L_1, D_1)$ are well separated from the eigenvalues of $(\widehat{L}_2, \widehat{D}_2)$, where the separation is measured by $\delta$, then the difference between the eigenspaces $\mathcal{R}(Z_1)$ and $\mathcal{R}(\widehat{Z}_1)$ is bounded as in (\ref{eigenspace_bound}). Moreover, for any simple eigenspace of $(L, D)$, the more separated the eigenvalue corresponding to this $1$-dimensional eigenspace is from $n-1$ eigenvalues of $(\widehat{L}_2, \widehat{D}_2)$ as measured by $\delta$ the more stable the eigenspace $\mathcal{R}(Z_1)$ is to perturbations in $L$ and $D$.
Similarly, the subspace spanned by the first $k$ eigenvectors of $(L, D)$ is stable to perturbations, if the corresponding $k$ eigenvalues of $(L_1,D_1)$ are well separated from $n-k$ eigenvalues of $(\widehat{L}_2, \widehat{D}_2)$.

\section*{Acknowledgment}
The authors would like to thank Prof. David Bindel for the helpful conversation on Matrix Perturbation Theory during the finalization of this work.

\bibliographystyle{plain}

\bibliography{zotero_references.bib}
\begin{IEEEbiography}{Gerald Ogbonna}
is a PhD candidate in the Systems Engineering department at Cornell University. He received a B.Sc. in Electrical and Electronics Engineering from Federal University of Technology Owerri, Nigeria and holds a M.S. in Systems Engineering from Cornell University. His research explores power network dynamics through the framework of coupled oscillator models, as part of the broader problem of control of networked multi-agent systems.
\end{IEEEbiography}

\begin{IEEEbiography}{C. Lindsay Anderson}
 is Professor and Chair of Biological and Environmental Engineering at Cornell University, with research affiliations in the Center for Applied Mathematics, Systems Engineering, and Electrical and Computer Engineering. Previously, she served as the Kathy Dwyer Marble and Curt Marble Faculty Director at the Cornell Atkinson Center for Sustainability and as interim Director of the Cornell Energy Systems Institute.  Research interests are the application of optimization under uncertainty to large-scale problems in sustainable energy systems. The National Science Foundation, US Department of Energy, US Department of Agriculture, PSERC, and the National Science and Engineering Research Council of Canada have supported her work. 
\end{IEEEbiography}
\end{document}